\documentclass[sigconf,natbib,screen=true]{acmart}

\AtBeginDocument{%
  \providecommand\BibTeX{{%
    \normalfont B\kern-0.5em{\scshape i\kern-0.25em b}\kern-0.8em\TeX}}}

\PassOptionsToPackage{dvipsnames}{xcolor}

\usepackage{graphicx}
\usepackage{subfigure}
\usepackage{bm}
\usepackage{bbm}
\usepackage{xspace}
\usepackage{url}
\usepackage{multirow}
\usepackage{booktabs}
\usepackage{enumerate}
\usepackage[inline]{enumitem}
\usepackage[skip=0pt]{caption}
\usepackage{acronym}
\usepackage{balance}
\usepackage{mleftright}

\usepackage{algorithm}
\usepackage{algorithmicx}

\usepackage[moderate,lists=normal,leading=normal]{savetrees}

\usepackage{makecell}
\copyrightyear{2022}
\acmYear{2022}
\setcopyright{rightsretained}
\acmConference[WSDM '22]{Proceedings of the Fifteenth ACM
International Conference on Web Search and Data Mining}{February
21--25, 2022}{Tempe, AZ, USA}
\acmBooktitle{Proceedings of the Fifteenth ACM International Conference
on Web Search and Data Mining (WSDM '22), February 21--25, 2022,
Tempe, AZ, USA}
\acmDOI{10.1145/3488560.3498375}
\acmISBN{978-1-4503-9132-0/22/02}

\newcommand{\ie}{\emph{i.e.,}\xspace}

\newcommand{\eg}{\emph{e.g.,}\xspace}

\newcommand{\ignore}[1]{}

\definecolor{HarrieGreen}{RGB}{34,139,34}

\acrodef{Seq-Rec}{sequential recommendation}
\acrodef{RS}{recommender system}
\acrodef{DANCER}{DebiAsing in the dyNamiC scEnaRio}
\acrodef{EIB}{error-imputation-{\allowbreak}based model}
\acrodef{IPS}{inverse propensity scoring}
\acrodef{DR}{doubly robust}
\acrodef{SNIPS}{self-normalized inverse propensity scoring}
\acrodef{MNAR}{missing not at random}
\acrodef{Pop}{Popularity}
\acrodef{Avg}{Average}
\acrodef{MF}{matrix factorization}
\acrodef{BPR}{Bayesian personalized ranking}
\acrodef{TMF}{time-aware matrix factorization}
\acrodef{TTF}{time-aware tensor factorization}
\acrodef{TMTF}{time-aware matrix \& tensor factorization}
\acrodef{MC}{Markov chains}
\acrodef{MDP}{Markov decision process}
\acrodef{GT}{Ground Truth}
\acrodef{sim-GT}{simulated Ground Truth}
\acrodef{CF}{collaborative filtering}
\acrodef{RNN}{recurrent neural network}
\acrodef{GRU}{gated recurrent unit}
\acrodef{LSTM}{long short-term memory}
\acrodef{SP}{static user preference}
\acrodef{DP}{dynamic user preference}
\acrodef{SB}{static selection bias}
\acrodef{DB}{dynamic selection bias}
\acrodef{SPSB}{static user preference with static selection bias}
\acrodef{SPDB}{static user preference with dynamic selection bias}
\acrodef{DPSB}{dynamic user preference with static selection bias}
\acrodef{DPDB}{dynamic user preference with dynamic selection bias}

\acrodef{NLL}{Negative Log-Likelihood}
\acrodef{PPL}{Perplexity}

\acrodef{P}{Precision}
\acrodef{R}{Recall}
\acrodef{MAP}{Mean Average Precision}
\acrodef{MRR}{Mean Reciprocal Rank}
\acrodef{NDCG}{Normalized Discounted
cumulative gain}

\acrodef{MSE}{Mean Squared Error}
\acrodef{MAE}{Mean Absolute Error}
\acrodef{ACC}{Accuracy}

\newcommand{\mysubsubsection}[1]{\vspace*{1mm}\noindent\emph{#1.~}}

\author{Jin Huang}
\affiliation{%
	\institution{University of Amsterdam}
	\city{Amsterdam}
	\country{The Netherlands}
}
\email{j.huang2@uva.nl}

\author{Harrie Oosterhuis}
\orcid{0000-0002-0458-9233}
\affiliation{%
	\institution{Radboud University}
	\city{Nijmegen}
	\country{The Netherlands}
}
\email{harrie.oosterhuis@ru.nl}
 
\author{Maarten de Rijke}
\orcid{0000-0002-1086-0202}
\affiliation{
 \institution{University of Amsterdam}
 \city{Amsterdam}
\country{The Netherlands}
}
\email{m.derijke@uva.nl}

\acmSubmissionID{26}

\begin{document}

\title[Debiasing Recommendations When Selection Bias and User Preferences Are Dynamic]{It Is Different When Items Are Older: Debiasing Recommendations When Selection Bias and User Preferences Are Dynamic}

\begin{abstract}
User interactions with \acfp{RS} are affected by user selection bias, \eg users are more likely to rate popular items (popularity bias) or items that they expect to enjoy beforehand (positivity bias).
Methods exist for mitigating the effects of selection bias in user ratings on the evaluation and optimization of \acp{RS}.
However, these methods treat selection bias as static, despite the fact that the popularity of an item may change drastically over time and the fact that user preferences may also change over time.

We focus on the age of an item and its effect on selection bias and user preferences.
Our experimental analysis reveals that the rating behavior of users on the MovieLens dataset is better captured by methods that consider effects from the age of item on bias and preferences.
We theoretically show that in a dynamic scenario in which both the selection bias and user preferences are dynamic, existing debiasing methods are no longer unbiased.
To address this limitation, we introduce \ac{DANCER}, a novel debiasing method that extends the \acl{IPS} debiasing method to account for dynamic selection bias and user preferences.
Our experimental results indicate that \ac{DANCER} improves rating prediction performance compared to debiasing methods that incorrectly assume that selection bias is static in a dynamic scenario.
To the best of our knowledge, \ac{DANCER} is the first debiasing method that accounts for dynamic selection bias and user preferences in \acp{RS}.
\end{abstract}



\maketitle

\acresetall

\setlength{\abovedisplayskip}{3pt}
\setlength{\belowdisplayskip}{3pt}

\section{Introduction}
User interactions with \acp{RS} are subject to selection bias, as a consequence of the selective behavior of users and of the fact that \acp{RS} actively restrict the items from which a user can choose~\citep{schnabel2016recommendations, ovaisi2020correcting, marlin2009collaborative, pradel2012ranking,steck2011item}.
A typical form of selection bias in \acp{RS} is \emph{popularity bias}: popular items are often overrepresented in interaction logs because users are more likely to rate them~\citep{pradel2012ranking, steck2011item, canamares2018should}.
Without correction, bias can affect user preference prediction~\citep{huang2020keeping,schnabel2016recommendations,yao2021measuring} and lead to problems of over-specialization~\cite{adamopoulos2014over}, filter bubbles~\cite{nguyen2014exploring, pariser2011filter}, and unfairness~\cite{chen2020bias}.
To correct for selection bias in interaction data from \acp{RS}, the task of \emph{debiased recommendation} has been proposed.
A widely-adopted method for this task makes use of \ac{IPS}, a causal inference technique~\cite{imbens2015causal}, and integrates it in the learning process of rating-prediction for recommendation~\cite{schnabel2016recommendations, huang2020keeping, chen2019correcting, joachims2017unbiased}.
It estimates the probability of a rating to be observed in the dataset, and inversely weights ratings according to these probabilities so that in expectation each user-item pair is equally represented.

While the existing \ac{IPS}-based debiasing method improves recommendations over methods that ignore the effect of bias, we identify two significant limitations. 
The way that \ac{IPS}-based debiasing is being applied for recommendations assumes that
\begin{enumerate*}[label=(\arabic*)]
\item the effect of selection bias is static over time, and
\item user preferences remain unchanged as items get older.
\end{enumerate*}
As we will show in Section~\ref{sec:ignoreDB}, current \ac{IPS}-based methods are unable to debias recommendations when the selection bias and user preferences are dynamic, \ie when they change over time.

\begin{figure}[t]
    \centering
    \includegraphics[width=1\linewidth]{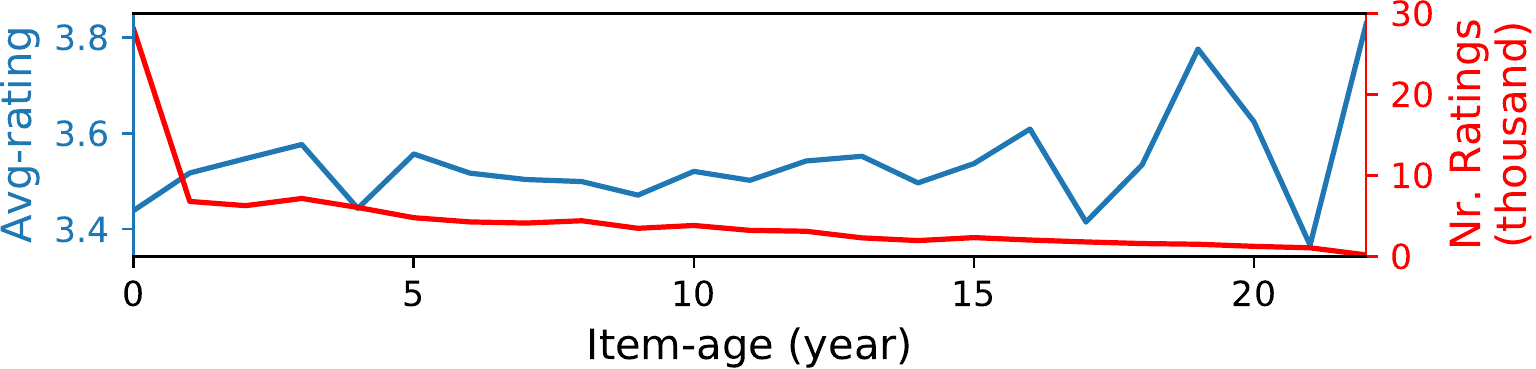}%
    \caption{The number of ratings (indicative of popularity) and the average (observed) rating of items for different item-ages on the MovieLens-Latest-small dataset.}
    \label{fig:DynamicChangedBias}
  \end{figure}

In practice, selection bias is usually dynamic, not static~\citep{ji2020re,chen2020bias}. 
Typically, the popularity of an item changes with item-age~\citep{chakraborty2017optimizing,ji2020re},
\ie the time since its publication.
Figure~\ref{fig:DynamicChangedBias} shows the number of ratings items received as they get older in the MovieLens dataset (red line).\footnote{\url{https://grouplens.org/datasets/movielens/latest/}}
On average, items receive the most attention during a short initial period of time after being published.
Hence, instead of \acl{SB}, real-world user behavior may be better captured with \emph{\acl{DB}} that assumes different probabilities of observing user ratings at different item-ages.
Besides selection bias, 
user preferences may also change over time~\citep{wangwatcharakul2020dynamic, al2017review, jagerman-2019-people}.
In this paper, we will focus on the effect of item-age on user preferences, and thus, on capturing the change in user preferences as items become older.
From Figure~\ref{fig:DynamicChangedBias}, it is clear that the average observed user rating varies with the item-age (blue line), despite the increased variance observed due to a decreasing number of logged interactions. 
We use the term \emph{dynamic scenario} to refer to the combination of \acl{DB} and \aclp{DP} occurring in a recommendation setting.

In this paper we first analyze real-world logged data to verify that the dynamic scenario is real: selection bias and user preferences are dynamic.
The dynamic scenario poses a two-fold problem for existing \ac{IPS}-based debiasing methods for \acp{RS}.
First, they are not unbiased in dynamic scenarios.
Second, existing methods~\citep{schnabel2016recommendations,canamares2018should} for estimating static selection bias cannot be used to estimate \acl{DB}.
Hence, we propose and evaluate a debiasing method to account for dynamic selection bias and dynamic user preferences.

All in all, we make a three-fold contribution:
\begin{enumerate*}[leftmargin=*,nosep]
    \item an analysis and estimation of \acl{DB} and \aclp{DP} in the MovieLens dataset;
    \item \acs{DANCER}: a general debiasing method that is adaptable for \acl{DANCER}; and
    \item \ac{TMF}-\acs{DANCER}: 
    to our knowledge it is the first recommendation method that corrects for \acl{DB} and models \aclp{DP}.  
\end{enumerate*}

\vspace*{-1mm}
\section{Related Work}

\emph{General Recommendation.~}
Early work on \acp{RS} typically uses collaborative filtering (\acs{CF}) to predict user ratings on items or make recommendations to users based on the feedback of similar users with similar behavior.
It is customary to divide recommendation tasks into the rating prediction task with explicit feedback (\eg user ratings) and the top-$K$ ranking task with implicit feedback (\eg clicks).
In this paper, we focus on rating prediction with explicit feedback.
The traditional \acf{MF} algorithm directly embeds users and items as vectors and models user-item interactions with an inner product~\citep{koren2009matrix,han2019auc}.
Some recent work has used deep neural networks to improve \acs{CF}, \eg by using multi-layer perceptrons~\citep{he2017neural,cheng2016wide}, convolutional neural networks~\citep{he2018outer}, or graph neural networks~\citep{he2020lightgcn, wang2020disentangled}.
While they significantly improve recommendation accuracy~\cite{koren2009bellkor}, they ignore the effect of time.

\mysubsubsection{Time-aware Recommendation}
Recently, a wide range of algorithms have been proposed that consider temporal information to improve \acp{RS}. 
Such methods are often classified as \emph{time-aware} or \emph{sequence-aware} recommendation methods.
Sequence-aware recommendation methods focus on the sequential order of interactions and aim to capture a user's short-term preferences~\citep{quadrana2018sequence}. Various deep learning methods have been applied to this task~\citep{quadrana2018sequence,zhao2021recbole} such as recurrent neural networks~\citep{wu2017recurrent,yu2016dynamic,hidasi2015session}, graph neural networks~\citep{xu2019graph,wu2019session}, and networks with attention~\citep{chen2018sequential,huang2018improving,sun2019bert4rec}.

We focus on time-aware recommendation methods~\citep{campos2014time} rather than sequence-aware recommendation methods, by considering changes in user preferences over exact time-periods.
One of the best known examples is \acf{TMF}~\citep{koren2009collaborative}, which takes the effect of time into consideration by adding time-dependent terms to the \ac{MF} model, thus allowing predicted ratings to vary over time.
\citet{koren2009collaborative} lists and compares various variants of \ac{TMF}, in how well they can capture item-related or user-related temporal effects.
\citet{xiong2010temporal} propose \ac{TTF}: a factorization based model that uses additional latent factors for each time period based on a probabilistic latent factor model.
Lastly, the effect of time is sometimes modelled by utilizing contextual attributes related to time (\eg day of the week or season of the year) as input features for context-aware \acp{RS}~\citep{baltrunas2009towards,panniello2009experimental,campos2014time,unger2020context}.

\mysubsubsection{Debiased Recommendation}
\label{sec:relatedwork:bias}
User selection bias is prevalent in logged data, meaning that many logged user ratings are \ac{MNAR}~\citep{hernandez2014probabilistic,marlin2009collaborative,schnabel2016recommendations}.
Two typical forms of bias in \acp{RS} are known as \emph{popularity bias} and \emph{positivity bias}. %
Popularity bias is characterized by a long tail distribution over the number of interactions per item in logged data because users are more likely to interact with more popular items~\citep{steck2011item, pradel2012ranking}.
Positivity bias leads to an over-representation of positive feedback because users rate the items they like more often~\citep{pradel2012ranking}.
The effect of these biases is generally dynamic: they can change drastically over time~\citep{ji2020re,chen2020bias,zhang2021causal}.
For instance, items are rarely popular for very extended periods of time, and therefore, we may expect a dynamic effect between the age of items and popularity bias.

Existing debiasing methods for reducing the effect of selection bias address \ac{MNAR} problems as follows:
\begin{enumerate*}
    \item the \ac{EIB} fills in missing ratings with predicted values, which may introduce bias due to inaccurate predictions~\citep{steck2010training},
    \item \acf{IPS} weights the loss associated with each observed rating inversely to their propensity, \ie the probability of observing that rating~\citep{schnabel2016recommendations, joachims2017unbiased, chen2019correcting}, and
    \item the \ac{DR} method integrates the \ac{EIB} and \ac{IPS} approaches to overcome the high variance of \ac{IPS} and the potential bias of \ac{EIB}~\citep{wang2019doubly}.
\end{enumerate*}

While the impact of dynamic bias has previously been pointed out~\citep{zhang2021causal,ji2020re}, no prior debiasing method considers a scenario in which both selection bias and user preferences change over time.
All existing debiased recommendation methods assume a static effect of selection bias regardless of whether they model dynamic user preferences.
Hence, there is currently no method that can effectively correct for bias in the dynamic scenario.
This is the research gap that we address.

\vspace*{-1.5mm}
\section{Problem Definition}
\label{sec:problem definition}

We follow the common \ac{RS} setting where items from the set $\mathcal{I} = \{i_1, \ldots, i_M\}$ are recommended to users from the set $\mathcal{U}=\{u_1, \ldots, u_N\}$~\citep{steck2013evaluation}.
Users have preferences towards items, generally modelled by a label $y_{u,i,t}$ (\eg a rating $y_{u,i,t} \in \{1,2,3,4,5\}$) per user $u \in \mathcal{U}$ and item $i \in \mathcal{I}$.
Similar to time-aware recommendations~\citep{campos2014time, koren2009collaborative, xiong2010temporal}, we also consider the effect of time on user preferences:
let $\mathcal{T} = \{t_1,\ldots,t_T\}$ be a set of $T$ time periods; we allow the user preference $y_{u,i,t}$ to vary over different periods $t \in \mathcal{T}$. 
Our goal is to optimize an \ac{RS} that best captures the user preferences across all items $i$ and time periods $t$.
We formulate this goal as a loss function: let $\hat{y}_{u,i,t}$ be a predicted rating by the \ac{RS} and $L(\hat{y}, y)$ a comparison function between the predicted rating and actual rating. Then our loss is:
\begin{equation}
	\mathcal{L} = \frac{1}{| \mathcal{U}|\cdot | \mathcal{I}|\cdot | \mathcal{T}|} \sum_{u\in \mathcal{U}}  \sum_{i\in \mathcal{I}} \sum_{t \in \mathcal{T}} L(\hat{y}_{u,i,t}, y_{u,i,t}).
	\label{eq:standard_loss}
\end{equation}
The function $L$ can be chosen according to common \ac{RS} metrics, for example, the prevalent \ac{MSE} metric:
\begin{equation}
L(\hat{y}_{u,i,t}, y_{u,i,t}) = (\hat{y}_{u,i,t} - y_{u,i,t})^2.
\end{equation}
The choice for \acp{RS} to perform well across all time periods $t$ in $\mathcal{T}$ is partially made for practical reasons; arguably, at any particular time one only needs \acp{RS} to perform well for the present and future~\citep{jagerman-2019-people}.
However, in practice, data is only available about past user preferences, thus making optimization w.r.t.\ future preferences infeasible.
Moreover, we expect that if \acp{RS}' performance generalizes well across the time periods in $\mathcal{T}$, it likely also generalizes well into the near future.

In our setting, logged interaction data is available to provide user ratings that can be used for optimization.
However, it is unrealistic for all users to provide ratings for all items. 
In practice user interaction data is very sparse.
We will use an observation indicator matrix $\mathbf{O}\in \{0,1\}^{|\mathcal{U}| \cdot |\mathcal{I}| \cdot |\mathcal{T}|}$ that indicates what ratings are recorded in the logged interaction data and during which time period.
We use $o_{u,i,t} \in \mathbf{O}$ to indicate this per rating: $o_{u,i,t} = 1$ indicates that the rating for user $u$ on item $i$ during time period $t$ has been recorded in the logged data, and $o_{u,i,t} = 0$ that it is missing.
The matrix $\mathbf{O}$ is strongly influenced by selection bias: certain ratings are much more likely to be observed than others.
This can be due to self-selection bias: users choosing to rate certain items more often~\citep{pradel2012ranking,steck2011item}; or algorithmic bias: the \ac{RS} used for logging choosing to show certain items more often~\citep{hajian2016algorithmic,baeza2018bias}.
Well-known prevalent biases in \ac{RS} data include:
\begin{enumerate*}[label=(\arabic*)]
\item popularity bias~\citep{steck2011item,pradel2012ranking} -- often a small group of popular items receive most interactions; and
\item positivity bias~\citep{pradel2012ranking} -- users are usually more likely to rate items they prefer.
\end{enumerate*}
We model selection bias using the probability of a rating being recorded: $p_{u,i,t} = P(o_{u,i,t} = 1)$, which we also refer to as the \emph{observation probability} or \emph{propensity}.
Again, we deviate from the common existing method by explicitly allowing $p_{u,i,t}$ to vary over different time periods $t$.
This enables our method to not only model a bias such as popularity bias but also how that bias changes as items get older and decline in popularity.

\vspace*{-1mm}
\section{Estimation Ignoring Dynamic Bias}
\label{sec:ignoreDB}

Before we introduce our recommendation method for dealing with the dynamic scenario in which both selection bias and user preferences are dynamic, 
we will show that, in a dynamic scenario, the existing recommendation methods
that either assume no bias or static bias are not unbiased.
The standard estimation of how well the predicted user preferences reflect the true user preferences shown in Eq.~\ref{eq:standard_loss} is the full-information loss (\ie the loss based on all the ratings), which is impractical since user preferences are only partially known in the logged data.
The naive loss ignores the effect of selection bias completely and thus assumes that the observed data represents the true user preferences unbiasedly.
Under this assumption, the naive loss can be estimated by a simple average on the observed ratings:
\begin{equation}
    \mathcal{L}_\text{Naive} = \frac{1}{|\{u,i,t : o_{u,i,t}=1\}|} \sum_{u,i,t : o_{u,i,t}=1} L(\hat{y}_{u,i,t}, y_{u,i,t}).
    \label{eq:naive_loss}
\end{equation}
And the widely-used debiasing method uses \ac{IPS} estimation~\cite{imbens2015causal, little2019statistical} to correct for the probability that a user rates an item~\citep{schnabel2016recommendations}.
It uses static propensities $p_{u,i}$ that are the probability of observing a rating for item $i$ by user $u$ in any of the time-periods~\citep{marlin2009collaborative, rosenbaum2002overt}.
These propensities ignore the dynamic aspect of selection bias, \ie that these probabilities can vary per time period $t$, resulting in the \emph{static} \ac{IPS} estimator:
\begin{equation}
    \mathcal{L}_\text{staticIPS} = \frac{1}{| \mathcal{U}|\cdot | \mathcal{I}|\cdot | \mathcal{T}|} \sum_{u,i,t : o_{u,i,t}=1} \frac{L(\hat{y}_{u,i,t}, y_{u,i,t})}{p_{u,i}}.
    \label{eq:staticips}
\end{equation}
Now that we have described the naive and \emph{static} IPS-based loss functions for recommendation (that assume no bias and only static bias, respectively), we can consider the effect of \acl{DB}.

\vspace*{-2mm}
\subsection{Effect of Dynamic Selection Bias}
Ignoring \acl{DB}, the recommendation methods 
that use the naive or \emph{static} IPS estimation are not unbiased in dynamic scenarios.
To illustrate how this may happen, we use a simple example $\mathcal{X}$ with one user $u$, one item $i$ and two time periods $t_1$ and $t_2$.
Let $y_{t_1}$ and $y_{t_2}$ be the user ratings on the item at $t_1$ and $t_2$ respectively; $p_{t_1}$ and $p_{t_2}$ denote the probabilities of observing the ratings at $t_1$ and $t_2$, respectively. 
We omit the subscript of $u$ and $i$ if no confusion can arise.
Due to \aclp{DP} and \acl{DB}, the user ratings and observation probabilities are not constant over the different time periods: 
$y_{t_1} \not= y_{t_2},\;\ \;\; p_{t_1} \not= p_{t_2}.$
Remember that in this example the loss we wish to estimate is:
\begin{equation}
    \mathcal{L}^\mathcal{X} = \frac{1}{2} \mleft( L(\hat{y}_{t_1}, y_{t_1})+ L(\hat{y}_{t_2}, y_{t_2}) \mright).
\end{equation}
The expected naive loss over the observation variables becomes:
\begin{equation}
\begin{split}
&\mathbb{E}\mleft[\mathcal{L}^\mathcal{X}_\text{Naive}\mright]  \\
&{}=p_{t_1}L(\hat{y}_{t_1}, y_{t_1}) + p_{t_2}L(\hat{y}_{t_2}, y_{t_2}) - \frac{p_{t_1}p_{t_2}}{2}\mleft(L(\hat{y}_{t_1}, y_{t_1})+ L(\hat{y}_{t_2}, y_{t_2})\mright).
\end{split}
\end{equation}
Clearly, it is not proportional to the true loss $\mathcal{L}^\mathcal{X}$ when selection bias and user preferences are dynamic: if $y_{t_1} \not= y_{t_2}$ and $p_{t_1}\not = p_{t_2}$, then $\mathbb{E}\mleft[\mathcal{L}^\mathcal{X}_\text{Naive}\mright] \not\propto \mathcal{L}^\mathcal{X}$.
This happens because the rating with the higher probability of being observed is over-represented in the observations.

Then the \emph{static} \ac{IPS}-based debiasing method uses static propensity $p_{u,i} = p_{t_1} + (1-p_{t_1})p_{t_2}$ that is the probability of observing a rating at time $t_1$ or $t_2$.
If we consider the expected value of this estimator:
\begin{equation}
    \begin{split}
    \mathbb{E}\mleft[\mathcal{L}^\mathcal{X}_\text{staticIPS}\mright]
    &= \frac{1}{2} \mleft(\frac{p_{t_1}}{p_{u,i}}L(\hat{y}_{t_1}, y_{t_1}) + \frac{p_{t_2}}{p_{u,i}}L(\hat{y}_{t_2}, y_{t_2}) \mright),
    \end{split}
\end{equation}
we see that it is not proportional to the true loss in the dynamic scenario: if $y_{t_1} \not= y_{t_2}$ and $p_{t_1}\not = p_{t_2}$, then  $\mathbb{E}\mleft[\mathcal{L}^\mathcal{X}_\text{staticIPS}\mright] \not\propto \mathcal{L}^\mathcal{X}$, 
because the \emph{static} IPS estimation fails to address the problem that the user's rating at a time with a higher probability of being observed is more likely to be represented in logged data than at any other time.
We note that the above counterexample holds regardless of whether the prediction of user ratings allows for dynamic preferences, \ie whether $\hat{y}_{t_1} = \hat{y}_{t_2}$ or $\hat{y}_{t_1} \not= \hat{y}_{t_2}$.

Our example is overly simplistic as it only contains a single user and a single item and two time periods; however, it can trivially be extended to any number of items, users or time periods.
Thus, it is a significant problem for \acp{RS} that optimization with the naive or \emph{static} \ac{IPS} is not unbiased if both the user preferences and the selection bias are dynamic; it will lead to biased optimization.
Selection bias and user preferences are practically never static in the real-world; in support of this claim,
Sections~\ref{sec:RQ1} and~\ref{sec:RQ2} provide evidence that the dynamic nature of bias and preferences can be observed in the MovieLens dataset.

\vspace*{-1mm}
\section{DANCER: Debiasing Recommendations in the Dynamic Scenario}
\label{sec:method}
We introduce \acs{DANCER}, a method for \acl{DANCER}.
We apply \acused{DANCER}\ac{DANCER} to \acf{TMF}, 
resulting in a novel rating prediction method that corrects for dynamic bias and models dynamic preferences.
We introduce a propensity estimation method to estimate the probabilities of ratings being observed per time period.

\vspace*{-2mm}
\subsection{Debiasing Recommendations}
As discussed in Section~\ref{sec:ignoreDB}, existing debiasing methods that use the naive or \emph{static} IPS estimation are unable to debias in the dynamic scenario where selection bias and user preferences are both dynamic.
As a solution, we propose \ac{DANCER}.
With accurate propensities $p_{u,i,t}$, \acl{DB} can be fully corrected by applying \ac{DANCER} to inversely weight the evaluation of the predicted ratings:
\begin{equation}
		\mathcal{L}_\text{\ac{DANCER}} = \frac{1}{| \mathcal{U}|\cdot | \mathcal{I}|\cdot | \mathcal{T}|} \sum_{u,i,t : o_{u,i,t}=1}  \frac{L(\hat{y}_{u,i,t}, y_{u,i,t})}{p_{u,i,t}}.
\end{equation}
Unlike the naive approach $\mathcal{L}_\text{Naive}$ (Eq.~\ref{eq:naive_loss}) and the \emph{static} IPS approach with a \emph{static} estimator $\mathcal{L}_{\text{staticIPS}}$ (Eq.~\ref{eq:staticips}), the proposed debiasing method $\mathcal{L}_{\text{\ac{DANCER}}}$ is unbiased in the dynamic scenario:
\begin{align}
        \mathbb{E}\mleft[\mathcal{L}_\text{\ac{DANCER}}\mright]
        &= \frac{1}{| \mathcal{U}|\cdot | \mathcal{I}|\cdot | \mathcal{T}|}  \sum_{u\in \mathcal{U}}  \sum_{i\in \mathcal{I}} \sum_{t \in \mathcal{T}} \frac{\mathbb{E}\mleft[o_{u,i,t}\mright] }{p_{u,i,t}} \cdot L(\hat{y}_{u,i,t}, y_{u,i,t})
        \nonumber \\
        &= \frac{1}{| \mathcal{U}|\cdot | \mathcal{I}|\cdot | \mathcal{T}|}  \sum_{u\in \mathcal{U}}  \sum_{i\in \mathcal{I}} \sum_{t \in \mathcal{T}} L(\hat{y}_{u,i,t}, y_{u,i,t})  \propto \mathcal{L}.
    \end{align}
Because \ac{DANCER} utilizes propensities that vary per time period $t$, it can correct for dynamic effects of bias that the existing static IPS estimators cannot.
For instance, in our example $\mathcal{X}$ with a user, an item and two time periods (see Section~\ref{sec:ignoreDB}), 
the expected \ac{DANCER} loss becomes:
\begin{equation}
    \begin{split}
    &\mathbb{E}\mleft[\mathcal{L}^\mathcal{X}_\text{DANCER}\mright] = \frac{1}{2} \mleft( p_{t_1} \frac{L(\hat{y}_{t_1}, y_{t_1})}{p_{t_1}} + p_{t_2}  \frac{L(\hat{y}_{t_2}, y_{t_2})}{p_{t_2}}\mright) = \mathcal{L}^\mathcal{X},
    \end{split}
\end{equation}
where we can see that $\mathcal{L}^\mathcal{X}_\text{DANCER}$ is an unbiased estimation of the true loss $\mathcal{L}^\mathcal{X}$.
Combined with a time-aware recommendation method, \ac{DANCER} is able to predict that the user ratings change over time.

\vspace*{-2mm}
\subsection{A Debiased Time-Aware Recommendation}

Because we expect both selection bias and user preferences to change over time in a dynamic scenario, the rating prediction that is optimized by \ac{DANCER} should also be able to account for changes in user preferences.
While \ac{DANCER} is not model specific, we will apply it to a \acf{TMF}~\cite{koren2009collaborative} model that accounts for temporal effects.
We refer to this combination of \ac{TMF} and debiasing method as \ac{TMF}-\ac{DANCER}.
Given an observed rating $y_{u,i,t}$ from user $u$ on item $i$ at time $t$, 
\ac{TMF} computes the predicted rating $\hat{y}_{u,i,t}$ as: 
$\hat{y}_{u,i,t}=\bm{p}_u^T\bm{q}_i+b_u+b_i+b+b_t$, 
where the $\bm{p}_u \in \mathbb{R}^{d}$ and $\bm{q}_i \in \mathbb{R}^{d}$ are embedding vectors of user $u$ and item $i$, and $b_u \in \mathbb{R}$, $b_i \in \mathbb{R}$, and $b \in \mathbb{R}$ are user, item and global offsets, respectively.
Crucially, $b_t$ is a time-dependent offset and models the impact of time in rating prediction.
Under this model, the proposed \ac{TMF}-\ac{DANCER} is optimized by minimizing the following loss:
\begin{equation}
    \label{eq-TMF_IPS}
    \mbox{}\hspace*{-1.5mm}
        \mathop{\arg~\,\min}_{\bm{P,Q,B}}
        \left[ 
        \sum\limits_{u,i,t:o_{u,i,t}=1}\frac{\delta(\hat{y}_{u,i,t}, y_{u,i,t})}{p_{u,i,t}}+\lambda\left(\|\bm{P}\|^2_F+\|\bm{Q}\|^2_F+\|\bm{B}\|^2_F \right)
        \right],
        \hspace*{-1mm}\mbox{}
\end{equation}
where $\bm{P}$, $\bm{Q}$ and $\bm{B}$ denote the embeddings of all users, all items and all the offset terms, respectively; $\delta$ is the \ac{MSE} loss function.

\vspace*{-2mm}
\subsection{Propensity Estimation}
\label{sec:propest}
\ac{DANCER} requires accurate propensities $p_{u,i,t}$ to remove the effect of \acl{DB}.
Because it is the first method to consider dynamic selection bias in \acp{RS}, it thus also needs a novel method to estimate $p_{u,i,t} = P(o_{u,i,t}=1)$, \ie the probability that the rating for user $u$ and item $i$ is observed at time $t$.
We propose to apply a \ac{NLL} loss to the propensity estimates $\hat{p}_{u,i,t}$ and the observations made in a dataset (indicated by $o_{u,i,t}$):
\begin{equation}
    \label{eq:OP_loss}
    \mathcal{L}_{\text{PE}} = 
    \frac{1}{| \mathcal{U}|\cdot | \mathcal{I}|\cdot | \mathcal{T}|}  \sum_{u\in \mathcal{U}}  \sum_{i\in \mathcal{I}} \sum_{t \in \mathcal{T}} L_{o}(\hat{p}_{u,i,t}, o_{u,i,t}),
\end{equation}
where the function $L_o$ is the \ac{NLL} for each individual propensity:
\begin{equation}
    \label{eq:nll}
    L_o(\hat{p}_{u,i,t}, o_{u,i,t}) = o_{u,i,t} \cdot \log~ \hat{p}_{u,i,t} + (1-o_{u,i,t}) \cdot \log(1-\hat{p}_{u,i,t}).
\end{equation}
Due to the large number of estimated propensities $\hat{p}_{u,i,t}$, we argue that it is best to predict them with a model.
Similar to the rating prediction task, \ac{TMF} and \ac{TTF}~\cite{xiong2010temporal} are potential choices to model how the propensities vary over users, items and time periods.
Alternatively, one can also make simplifying assumptions in the estimations of dynamic popularity bias. 
For instance, $\hat{p}_{u,i,t} = \text{Pop}(i, t) := \frac{\sum_{u'\in \mathcal{U}}o_{u',i,t}}{|\mathcal{U}|}$ uses the ratio of ratings received by item $i$ at time $t$.
The $\text{Pop}(i, t)$ estimate is easy to compute, but it does assume that there are no differences between users when it comes to providing ratings.

Finally, we note that our proposed propensity estimation method Eq.~\ref{eq:OP_loss} builds on existing methods for propensity estimation for \acl{SB}.
\citet{saito2020unbiased} use \ac{MF} instead of \ac{TMF} or \ac{TTF}.
Similarly, the $\text{Pop}(i) := \frac{\sum_{u'\in \mathcal{U}} \sum_{t' \in \mathcal{T}} o_{u',i,t'}}{|\mathcal{U}|\cdot |\mathcal{T}|}$ is a common way to measure (static) popularity bias~\cite{zhang2021causal,ciampaglia2018algorithmic,canamares2018should}.
Our propensity estimation method makes these methods applicable to the dynamic scenario, and enables them to provide propensities for the \ac{DANCER} debiasing method.

\vspace*{-1mm}
\section{Experiments}
\label{sec:experiments}
In our experiments, we focus on the age of an item (item-age) and the dynamic effect it has on selection bias and user preferences.
From this point onwards, our notation will use $t$ to denote how long an item has been available in the system, we will refer to this as the \emph{age of the item}.

Because the distribution of ratings is very skewed towards young items, we divide the item-ages into seven bins whose edges are $[0, 1, 3, 5, 8, 11, 15, \infty]$ in years.
For instance, a rating on an item when it is two-and-a-half years old will be assigned to $t=2$, and a rating when it is 15 years old will be assigned $t=7$.
This can be interpreted as a specific choice for the time periods $\mathcal{T}$ and thus does not change any of the previously stated theory.

We first wish to investigate whether real-world selection bias and user preferences are affected by item-age -- and are thus dynamic -- and whether \ac{TMF}-\ac{DANCER} is more effective in a dynamic scenario than existing rating prediction methods that do not consider dynamic bias.
Our experimental analysis is organized around three research questions:
\begin{enumerate*}[label=(\textbf{RQ\arabic*})] 
	\item Does item-age affect selection bias present in logged data?
	\item Does item-age affect real-world user preferences?
	\item Does the proposed \ac{TMF}-\ac{DANCER} method better mitigate the effect of bias in the dynamic scenario than existing debiasing methods designed for static selection bias?
\end{enumerate*}
To answer these questions, we make use of three different tasks based around the
MovieLens-Latest-small dataset~\cite{harper2015movielens}.
The following sections will each introduce one of these tasks and answer the corresponding research question.

All tasks use embeddings with 32 dimensions, hyperparameter tuning is applied per method and task in the following ranges: 
learning rate $\eta \in \{10^{-5}, \ldots, 0.1\}$ and $L_2$ regularization weights $\lambda \in \{0, 10^{-7}, 10^{-6}, \ldots, 1.0\}$.
Our implementation and hyperparameter choices are available at \url{https://github.com/BetsyHJ/DANCER}.

\vspace*{-1mm}
\section{RQ1: Is Selection Bias Dynamic?}
\label{sec:RQ1}

To answer 
\textbf{RQ1}: \emph{Does item-age affect selection bias present in real-world logged data?}, 
we will evaluate whether methods that consider item-age can better predict which items will be rated than methods that do not.
If item-age has a large effect on selection bias, it should be an essential feature for predicting whether users will rate an item.

\vspace*{-2mm}
\subsection{Experimental Setup for RQ1}
\label{sec:models}

The goal of our first task is to predict which ratings will be observed in real-world data, in other words, across users $u$, items $i$ and item-ages $t$ the aim is to predict the observation $o_{u,i,t}$ variables.
With $\hat{p}_{u,i,t}$ as the predicted probability of observation, the metrics for this task are the \ac{NLL} (Eq.~\ref{eq:nll}) and \ac{PPL}: 
\begin{equation}
2^{-\frac{1}{| \mathcal{U}|\cdot | \mathcal{I}|\cdot | \mathcal{T}|}  \sum_{u\in \mathcal{U}}  \sum_{i\in \mathcal{I}} \sum_{t \in \mathcal{T}} o_{u,i,t} \cdot \log_2 \hat{p}_{u,i,t} + (1-o_{u,i,t}) \cdot \log_2(1-\hat{p}_{u,i,t})}.
\end{equation}
To evaluate whether item-age has a significant effect on the observation probabilities -- and thus the dynamic selection bias in the data --, we compare the performance of observation prediction methods that assume static bias with others that take item-age into account.
Our comparison contains three baselines, one static method and four time-aware methods; when specifying the methods, we use $\sigma$ to denote the sigmoid function, $\bm{p}_u$ for a learned user embedding, $\bm{q}_i$ for an item embedding, $\bm{a}_t$ for an embedding representing an item-age, and $b_t$ is a learned parameter that varies per item-age $t$. 
\begin{enumerate}[leftmargin=*, label=(\arabic*),nosep]
	\item \textbf{Constant}: The fraction of all ratings, this assumes no selection bias is present: $\hat{p}_{u,i,t} = \frac{\sum_{u'\in \mathcal{U}}  \sum_{i'\in \mathcal{I}} \sum_{t' \in \mathcal{T}} o_{u',i',t'}}{|\mathcal{U}|\cdot |\mathcal{I}|\cdot |\mathcal{T}|}$.
	\item \textbf{Static Item \ac{Pop}}: The fraction of all ratings that have been given to the item; this assumes that selection bias is static over users and time: $\hat{p}_{u,i,t} = \frac{\sum_{u'\in \mathcal{U}} \sum_{t' \in \mathcal{T}} o_{u',i,t'}}{|\mathcal{U}|\cdot |\mathcal{T}|}$.
	\item \textbf{Time-aware Item Popularity (T-\ac{Pop})}: 
	The item popularity per item-age; defined as the fraction of all ratings that have been given to item $i$ of age $t$: $\hat{p}_{u,i,t} = \frac{\sum_{u'\in \mathcal{U}}o_{u',i,t}}{|\mathcal{U}|}.$
	\item \textbf{Static \acf{MF}}: A standard \ac{MF} model that assumes selection bias is static: $\hat{p}_{u,i,t} = \sigma (\bm{p}_u^T\bm{q}_i)$.

	\item \textbf{\Acf{TMF}}~\cite{koren2009collaborative}: \ac{TMF} captures the drift in popularity as items get older by adding an age-dependent bias term: $\hat{p}_{u,i,t} = \sigma (\bm{p}_u^T\bm{q}_i + b_t)$.
	\item \textbf{\Acf{TTF}}~\cite{xiong2010temporal}: \ac{TTF} extends \ac{MF} by modelling the effect of item-age via element-wise multiplication: $\hat{p}_{u,i,t} = \sigma (\bm{p}_u^T(\bm{q}_i \times \bm{a}_t))$. 
	\item \textbf{\acs{TTF}++}: We propose a variation on \ac{TTF} that models the effect via summation instead: $\hat{p}_{u,i,t} = \sigma (\bm{p}_u^T(\bm{q}_i + \bm{a}_t))$.
	\item \textbf{\Acf{TMTF}}: Lastly, we propose a novel integration of \ac{TMF} with \ac{TTF}++: $\hat{p}_{u,i,t} = \sigma (\bm{p}_u^T(\bm{q}_i + \bm{a}_t) + b_t)$.
\end{enumerate}
All models are optimized with the \ac{NLL} loss as described in Section~\ref{sec:propest}.

We split the dataset into training, validation and test partitions following a ratio of 7:1:2.
The MovieLens-Latest-small dataset~\citep{harper2015movielens} consists of 100,836 ratings applied to 9,742 movies by 610 users between 1996 and 2018.
We apply two splitting strategies to the data:
\begin{enumerate*}[label=(\arabic*)]
\item a time-based split that per user places the latest 20\% of their ratings into the test set~\cite{campos2014time}; and
\item a random split that uniformly samples 20\% of ratings per user.
\end{enumerate*}
The time-based split is more realistic but makes the training and test data follow different distributions: \ie there will be more ratings on younger items in the training set than in the test set.
Alternatively, the random split ensures both partitions follow the same distribution but is less realistic: \ie ratings in the test set may have taken place before ratings in the training set.
For both settings, the training and validation set are uniformly randomly sampled from the data outside the test set.
Since most users have an active lifecycle of less than one year, the time-based split results in a ratio between observed and missing ratings that is four times higher than the ratio in the test set;
to account for this large difference in distributions we scale the predicted $\hat{p}_{u,i,t}$ by 0.25 in this setting.
This leads to considerable performance improvements for all methods.
Lastly, we ignore ratings outside of the user's presence in the dataset, \ie before their first rating or after their last; this prevents the methods from having to predict when users became active so that they can focus on the effect of item-age.

\begin{table}[tbp]
	\centering
	\setlength{\tabcolsep}{0.5mm}	
        \caption{RQ1 -- Performance in observation prediction. Results are averages of 10 independent runs, the standard deviations are shown in brackets. $\dag$ indicates a significant improvement over \ac{MF} ($p < 0.01$) according to the paired-samples t-test.} \vspace{1mm}
		\label{tab:OIPT}%
			\begin{tabular}{@{}l@{} l@{}l l@{}l l@{}l l@{}l@{}}
			\toprule
			\multirow{2}[1]{*}{Method}&\multicolumn{4}{c}{\textsc{Random}} &\multicolumn{4}{c}{\textsc{Time-Based}}
			\\ \cmidrule(lr){2-5} \cmidrule(lr){6-9}
			& NLL && PPL && NLL && PPL & \\
			\midrule
			Constant \  & 0.0973 & & 1.1022 & & 0.0337 & & 1.0343 & \\
			Pop & 0.0890 & & 1.0931 & & 0.0404 & & 1.0412 & \\
            MF & 0.0697&{\scriptsize(0.0015)} & 1.0722&{\scriptsize(0.0016)} & 0.0271&{\scriptsize(0.0000)} & 1.0275&{\scriptsize(0.0000)} \\
            \midrule
			T-Pop & 0.1234 & & 1.1314 & & 0.0523 & & 1.0537 & \\
            TMF & 0.0658$^\dag$ & {\scriptsize(0.0001)} & 1.0680$^\dag$ & {\scriptsize(0.0001)} & \textbf{0.0267}$^\dag$ & {\scriptsize(0.0000)} & \textbf{1.0271}$^\dag$ & {\scriptsize(0.0000)} \\
            TTF & 0.0637$^\dag$ & {\scriptsize(0.0002)} & 1.0657$^\dag$ & {\scriptsize(0.0003)} & 0.0273 & {\scriptsize(0.0004)} & 1.0277 & {\scriptsize(0.0004)} \\
            TTF++ & 0.0632$^\dag$ & {\scriptsize(0.0002)} & 1.0653$^\dag$ & {\scriptsize(0.0002)} & 0.0268$^\dag$ & {\scriptsize(0.0001)} & 1.0271$^\dag$ & {\scriptsize(0.0001)} \\
            TMTF & \textbf{0.0621}$^\dag$ & {\scriptsize(0.0001)} & \textbf{1.0641}$^\dag$ & {\scriptsize(0.0001)} & 0.0268$^\dag$ & {\scriptsize(0.0000)} & 1.0272$^\dag$ & {\scriptsize(0.0000)} \\
			\bottomrule
		\end{tabular}%
\end{table}%

\vspace*{-2mm}
\subsection{Results for RQ1}
The results for the first task are presented in Table~\ref{tab:OIPT}.
Clearly, under both splitting strategies, the time-aware methods \ac{TMF}, \ac{TTF}++ and \ac{TMTF} are significantly more accurate than \ac{Pop} and \ac{MF}, which assume that selection bias is static, while \ac{MF} outperforms Constant, which assumes no bias.
Interestingly, T-Pop performs worst among all the methods, probably due to the high variance caused by sparsity.

Under the random splitting strategy, \ac{TTF} and \ac{TTF}++ outperform \ac{TMF}, while \ac{TMTF} outperforms all other methods.
Thus it appears that modelling item-age via a learned embedding better captures its effect than a single learned parameter, but moreover, \ac{TMTF} shows us that combining both results in the most accurate method.
Under the time-based splitting strategy, \ac{TMF} performs slightly better than \ac{TTF}++ and \ac{TMTF}, while \ac{TTF} performs worse than them.
Also, \ac{Pop} performs worse than Constant.
A plausible reason for this inconsistency is the difference in distribution between the training and test set caused by the time-based split.
The number of ratings per year displayed in Figure~\ref{fig:avgrating-timebasedsplitting} displays this difference.
This suggests that \ac{TMF} is more robust to differences in distribution and that the other methods are somewhat overfitted on the training set.
Nevertheless, most time-aware methods still predict the selection bias significantly better than the static \ac{MF}.

We thus conclude that time-aware methods can better predict selection bias in real-world data than static methods.
While the skewed rating distribution 
in Figure~\ref{fig:DynamicChangedBias} already suggests that item-age has a large influence, our experimental results strongly show that item-age is an essential factor for accurately capturing the selection bias in users' rating behavior.
Consequently, we answer \textbf{RQ1} affirmatively: item-age significantly affects the selection bias present in real-world data.
This result strongly implies that the assumption of static bias in previous work is incorrect, at least in recommendation settings similar to that of the MovieLens dataset.

\vspace*{-1mm}
\section{RQ2: Are User Preferences Dynamic?}
\label{sec:RQ2}
To answer \textbf{RQ2}: \emph{Does item-age affect real-world user preferences?}, 
we compare rating prediction methods that assume preferences are static with ones that allow for dynamic preferences.
If item-age has a significant effect, the latter group should perform better.

\vspace*{-2mm}
\subsection{Experimental Setup for RQ2}

The average rating per item-age in Figure~\ref{fig:DynamicChangedBias} does not reveal a clear influence from the item-age on rating behavior.
However, the averages should not be taken at face value because they are subject to selection bias.
Users are generally more likely to rate movies they like (\ie positivity bias~\citep{pradel2012ranking}), thus it is possible that while the \emph{true} average rating drops, the \emph{observed} remains stable due to selection bias.

To find out whether item-age has a substantial effect, we compare methods that assume static preferences with others that allow for dynamic preferences in terms of the \acf{MSE}, \ac{MAE} and \ac{ACC} metrics.
We train and evaluate in two settings:
\begin{enumerate*}[label=(\arabic*)]
\item in the \emph{observed setting} the dataset is used without any corrections to mitigate selection bias; and
\item in the \emph{debiased setting} \ac{SNIPS}~\cite{swaminathan2015self, yang2018unbiased} is applied during training and metric calculation to mitigate the effect of selection bias.
\end{enumerate*}
The advantage of the debiased setting is that -- in expectation -- it bases evaluation on the true rating distribution; however, it has drawbacks: it requires accurate propensities and can be subject to increased variance.
The observed setting will provide biased estimates but does not have these drawbacks.
Our evaluation considers both settings so that their advantages can complement each other.

The comparison includes two baselines:
\begin{enumerate}[label=(\arabic*),leftmargin=*,nosep]
\item \textbf{Static Average Item Rating (Avg)}:
The average observed rating across all item-ages:
$\hat{y}_{u,i,t}=\frac{\sum_{u',i,t' : o_{u',i,t'}=1} y_{u',i,t'}}{\sum_{u'\in \mathcal{U}} \sum_{t' \in \mathcal{T}} o_{u',i,t'}}$.
\item \textbf{Time-aware Average Item Rating (T-Avg)}:
the average observed rating per item-age:
$\hat{y}_{u,i,t} = \frac{\sum_{u',i,t : o_{u',i,t}=1} y_{u',i,t}}{\sum_{u'\in \mathcal{U}}  o_{u',i,t}}$.
\end{enumerate}
In addition, we also compare with the static \ac{MF} and the time-aware \ac{TMF}, \ac{TTF}, \ac{TTF}++ and \ac{TMTF}.
These methods are analogous to those used in Section~\ref{sec:RQ1}; the main difference is that for this task the $\sigma$ sigmoid function is not applied.
Additionally,  we add a global offset $b$, a user offset $b_u$, and an item offset $b_i$ to \ac{MF}, \ac{TMF} and \ac{TMTF}.
All methods are optimized to minimize \ac{MSE}; in the debiased setting optimization is performed with \ac{DANCER} following Section~\ref{sec:method}.
We use the propensity values estimated for the previous observation prediction task by \ac{TMTF} under the random-split (see Section~\ref{sec:models}).

\begin{figure}[t]
	\centering
\includegraphics[width=0.50\linewidth, trim=0 0pt 0 20pt]{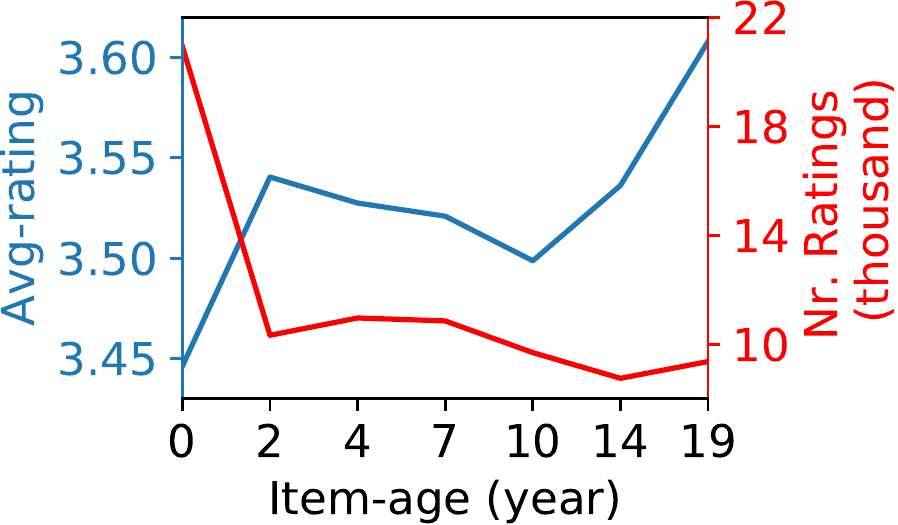}%
\includegraphics[width=0.50\linewidth, trim=0 0pt 0 20pt]{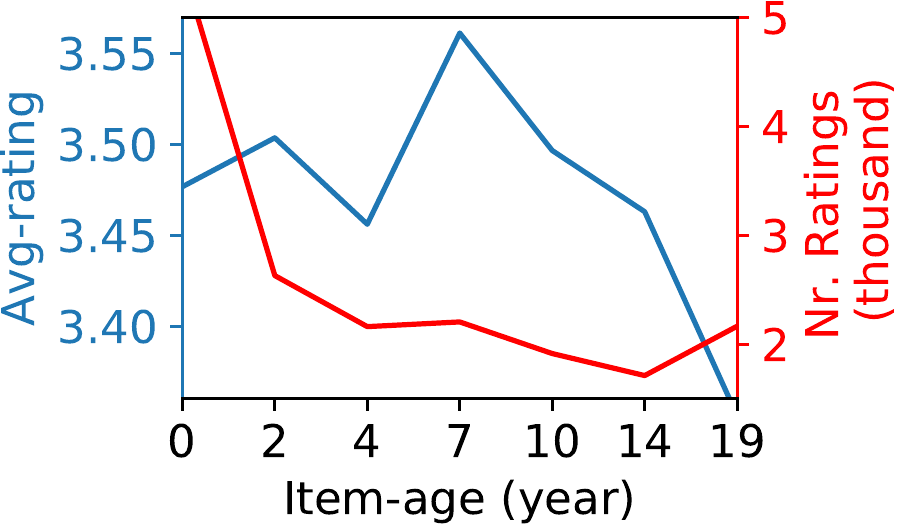}
    \caption{\fontdimen2\font=0.32ex Average rating and number of ratings over item-age in the time-based partitioned training (left) and test set (right). \looseness=-1
	}
	\vspace{-1mm}
	\label{fig:avgrating-timebasedsplitting} %
\end{figure}

The dataset is again partitioned into a training, validation and test set according to a ratio of 7:1:2.
Unlike for the previous task (Section~\ref{sec:models}), the data for this task only consists of observed ratings, and furthermore, the partitioning is only made via uniform random sampling.
As displayed in Figure~\ref{fig:avgrating-timebasedsplitting}, we find that a time-based split leads to extremely different rating distributions.
This makes it infeasible to obtain convincing conclusions from the results of this task.
Nevertheless, because a random split is perfectly suitable for evaluating a possible relationship between user preferences and item-age, our results are completely appropriate to answer \textbf{RQ2}.

\begin{table*}[tbp]
	\centering
        \caption{RQ2 -- Performance comparison of different methods in predicting ratings logged in MovieLens-Latest-small. 
        $\dag$ indicates that the improvement of the models over \ac{MF} is significant ($p < 0.01$).
        $\uparrow / \downarrow$ indicates whether larger or smaller values are better.} %
		\label{tab:OPPT}%
			\begin{tabular}{l l@{~}l l@{~}l l@{~}l l@{~}l l@{~}l l@{~}l}
			\toprule
			\multirow{2}[1]{*}{Method}&\multicolumn{6}{c}{\textsc{Observed}} &\multicolumn{6}{c}{\textsc{Debiased}}
			\\ \cmidrule(lr){2-7} \cmidrule(lr){8-13}
			& \multicolumn{2}{l}{MSE $\downarrow$} & \multicolumn{2}{l}{MAE  $\downarrow$} & \multicolumn{2}{l}{ACC$\uparrow $} & \multicolumn{2}{l}{SNIPS-MSE$\downarrow$} & \multicolumn{2}{l}{SNIPS-MAE$\downarrow$} & \multicolumn{2}{l}{SNIPS-ACC$\uparrow $} \\
			\midrule
			Avg & 0.9535 & & 0.7540 & & 0.2241 & & 1.1436 & & 0.8360 & & 0.2048 \\
            MF & 0.7551& {\scriptsize(0.0046)} & 0.6679 & {\scriptsize(0.0021)} & 0.2515 & {\scriptsize(0.0016)} & 1.2911 & {\scriptsize(0.0242)} & 0.8985 & {\scriptsize(0.0095)} & 0.1829 & {\scriptsize(0.0065)}  \\
            \midrule
			T-Avg & 1.0850 & & 0.7974 & & 0.2181 & & 1.3105 & & 0.8865 & & 0.1955 & \\
            TMF & 0.7505 & {\scriptsize(0.0058)} & 0.6656 & {\scriptsize(0.0026)} & 0.2525 & {\scriptsize(0.0014)}  & 1.1210$^\dag$ & {\scriptsize(0.0464)} & 0.8383$^\dag$ & {\scriptsize(0.0173)} & 0.1944$^\dag$ & {\scriptsize(0.0067)}  \\
            TTF & 1.1515 & {\scriptsize(0.0542)} & 0.8187 & {\scriptsize(0.0181)} & 0.2120 & {\scriptsize(0.0054)} & 1.8834 & {\scriptsize(0.1247)} & 1.0879 & {\scriptsize(0.0388)} & 0.1504 & {\scriptsize(0.0058)}  \\
            TTF++ & 0.7526 & {\scriptsize(0.0011)} & 0.6645$^\dag$ & {\scriptsize(0.0006)} & \textbf{0.2552}$^\dag$ & {\scriptsize(0.0007)}  & 1.0839$^\dag$ & {\scriptsize(0.0159)} & 0.8067$^\dag$ & {\scriptsize(0.0067)} & \textbf{0.2134}$^\dag$& {\scriptsize(0.0059)} \\
            TMTF & \textbf{0.7503}$^\dag$ & {\scriptsize(0.0014)} & \textbf{0.6637}$^\dag$ & {\scriptsize(0.0008)} & 0.2533$^\dag$ & {\scriptsize(0.0009)}  & \textbf{1.0727}$^\dag$ & {\scriptsize(0.0173)} & \textbf{0.8026}$^\dag$ & {\scriptsize(0.0047)} & 0.2127$^\dag$ & {\scriptsize(0.0060)} \\
			\bottomrule
		\end{tabular}%
	\vspace{-3mm}
\end{table*}%

\vspace*{-2mm}
\subsection{Results for RQ2}
Table~\ref{tab:OPPT} displays the evaluation results for the second task; in both settings the time-aware methods outperform the static \ac{MF}.
There is a single exception: \ac{TTF} performs worst in both settings, probably due to over-fitting. 
The differences between the other time-aware methods and static MF are larger in the debiased setting than in the observed setting.
This suggests that selection bias in the data reduces the dynamic effect of item-age on the observed ratings.
We speculate that the effect of positivity bias could increase with item-age: users are less likely to try and rate movies that are older unless they already expect to enjoy them.
Due to sparsity, T-Avg performs worse than Avg in both settings.
Interestingly, Avg performs even better than \ac{MF} in the debiased setting; this confirms prior observations that Avg is more robust in highly biased scenarios~\cite{canamares2018should}.
Regardless, in both settings most time-aware methods significantly outperform MF and the two baselines,
and therefore, we answer \textbf{RQ2} in the affirmative: item-age has a significant effect on user preferences.

Our conclusions for RQ1 and RQ2 indicate that the dynamic scenario, where selection bias and user preferences change over time, better captures real-world logged data, than a static view.
Moreover, Section~\ref{sec:ignoreDB} showed that the existing \emph{static} IPS approach cannot debias in this scenario.
Consequently, our answers to RQ1 and RQ2 reveal a real need for a method that can deal with the dynamic scenario.

\vspace*{-1mm}
\section{RQ3: Can \ac{TMF}-\ac{DANCER} Better Mitigate Dynamic Selection Bias?}
\label{sec:RQ3}
Section~\ref{sec:ignoreDB} showed that the \emph{static} \ac{IPS}-based debiasing method is biased in a dynamic scenario. Subsequently, in Section~\ref{sec:RQ1} and \ref{sec:RQ2} we discovered that selection bias and user preferences in the MovieLens dataset are indeed dynamic.
Therefore, we can already conclude that \emph{theoretically} \ac{TMF}-\ac{DANCER} is the first method that is potentially unbiased for the dynamic scenario.
Our final research question considers whether this theoretical advantage translates into improved recommendation performance: \textbf{RQ3}: \emph{Does the proposed \ac{TMF}-\ac{DANCER} method better mitigate the effect of bias in the dynamic scenario than existing debiasing methods designed for static selection bias?}

\vspace*{-2mm}
\subsection{Experimental Setup for RQ3}
The most common technique for evaluating debiasing methods for recommendation, without actual deployment to real-world users, makes use of unbiased test sets~\cite{schnabel2016recommendations, wang2019doubly}.
This requires a dataset that has a training set consisting of biased logged ratings and a test set of user ratings on uniformly randomly selected items.
Such a test set can be created by randomly sampling items and asking users to provide a rating for them, thus avoiding the selection bias that usually heavily affects what items are rated.
However, the publicly available datasets that meet this criterion -- \textsc{Yahoo!R3}~\cite{marlin2009collaborative} and \textsc{Coat Shopping}~\cite{schnabel2016recommendations} -- lack any form of temporal information.\footnote{A recent music dataset~\cite{brost2019music} contains randomized observations and temporal information, but it only tracks user behavior during short sessions rather than for extended periods of time.}
As a result, we cannot apply \ac{DANCER} or any other form of dynamic debiasing to them.

As an alternative to using real-world datasets, we utilize a semi-synthetic simulation based on a real-world dataset for our evaluation.
This simulation first estimates a \ac{sim-GT} based on the actual dataset, and then generates a new biased training set from this \ac{sim-GT}.
Debiasing methods can be applied to the generated training set and evaluated on the \ac{sim-GT}, since in this setting, the debiased estimates should match the \ac{sim-GT} as close as possible.
The creation of our semi-synthetic simulation has three steps:
\begin{enumerate}[label=(\arabic*),leftmargin=*,nosep]
	\item
	First, we estimate the complete rating matrix using the \ac{TMF} method,
	which simply uses an age-dependent bias term to model the dynamics of user preferences, thus making the simulation understandable and not prone to overfitting.
	This provides us with an estimated rating for each item, user and item-age combination which we will treat as the \ac{sim-GT}.
	By optimizing \ac{TMF} with the real user ratings in the debiased setting, we hope the \ac{sim-GT} reflects the real-world scenario as closely as possible.
	\item
	Second, dynamic selection bias is simulated using \ac{MF} to model the interactions between items and item-ages.
	Following Section~\ref{sec:RQ1}, we fit the following model: $p_{u,i,t} = \sigma(\bm{q}_i^T \bm{a}_t)$, to predict if the ratings are observed in the MovieLens dataset.
	To mimic real-world dynamic popularity bias more closely, we follow the user presence of the original dataset: propensities are zero before a user's first rating and after their last rating in the dataset, we also normalize the predicted probabilities so that their mean value is 4\%, the same value as the dataset has.
	\item
	Third, to prevent overlap between the training and test set, we utilize both random and time-based splitting:
	per user, 50\% of items are randomly selected for the test set, and a split timestep is chosen at 80\% of the user presence.
	The test set consists of all \ac{sim-GT} ratings on the randomly selected items at the last presence of each user;
	as a result, the test set reflects future preferences on previously unseen items.
	The training set uses the other 50\% of items per user and samples from the ratings before the split timestamp following the estimated propensities $p_{u,i,t}$ from the previous step.
	The result is a training set where due to dynamic selection bias only $\sim$2\% of the $y_{u,i,t}$ ratings are observed.
\end{enumerate}

\begin{figure}[t]
	\centering
	\includegraphics[width=0.725\linewidth]{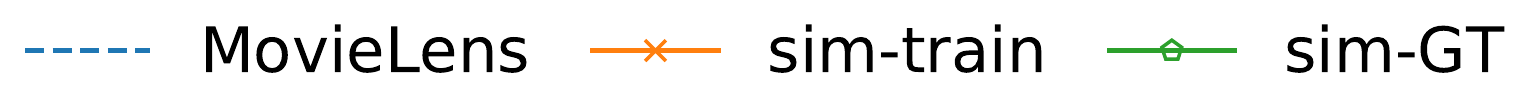}\\ \vskip -3pt
	\includegraphics[width=0.5077\linewidth]{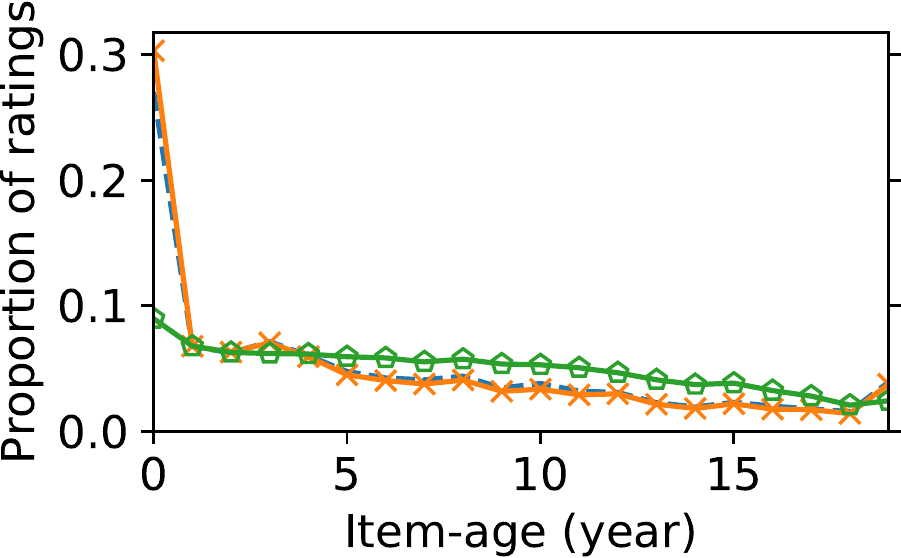}%
	\includegraphics[width=0.4923\linewidth]{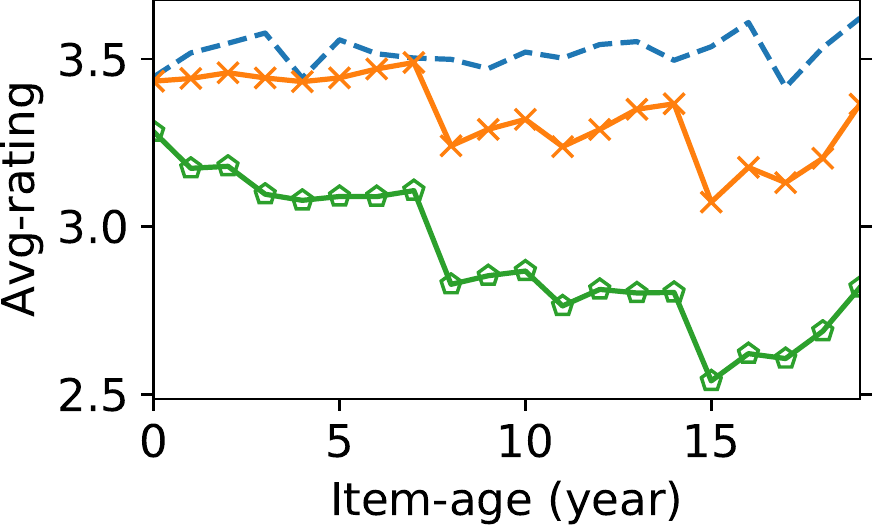}\\
    \caption{RQ3 -- The proportion of ratings and the average rating of items over item-age on MovieLens, the simulated training set (sim-train) and the \acf{sim-GT}.}
	\label{fig:simulated_data} %
\end{figure}

\noindent%
Figure~\ref{fig:simulated_data} compares the original MovieLens dataset with our semi-synthetic simulation.
The popularity of items, in terms of how many ratings they receive, is closely approximated by the simulated training set.
In terms of average rating, there is some deviation from the simulated training set and MovieLens: the simulated training set rates older items lower than MovieLens.
It seems likely that this is the result of positivity bias, which is not part of our simulation.
Nonetheless, we clearly see that both dynamic selection bias and dynamic user preferences are represented in our simulation.

We compare the performance of \ac{TMF}-\ac{DANCER} with the following baselines:
\begin{enumerate*}[label=(\arabic*)]
	\item Four methods that ignore bias altogether: Avg, T-Avg, \ac{MF} and \ac{TMF} (see Section~\ref{sec:RQ2}). %
	\item Two methods optimized with the \emph{static} IPS estimator: \ac{MF}-staticIPS~\cite{schnabel2016recommendations} and \ac{TMF}-staticIPS, which use the Static Item \acl{Pop} propensities from Section~\ref{sec:RQ1}.
	\item
	A static preference method with dynamic debiasing: \ac{MF}-\ac{DANCER}, which optimizes a (static) \ac{MF} while correcting for the effect of dynamic bias.
\end{enumerate*}

Finally, to evaluate whether \ac{TMF}-\ac{DANCER} is robust to misspecified propensities, we compare its performance with using Time-Aware General Popularity (TG-Pop):
$\hat{p}_{u,i,t} = \frac{\sum_{u'\in \mathcal{U}}\sum_{i'\in \mathcal{I}} o_{u',i',t}}{|\mathcal{U}|\cdot |\mathcal{I}|}$,
and 
Time-aware Item Popularity (T-\ac{Pop}) (see Section~\ref{sec:models}).

\vspace*{-2mm}
\subsection{Results for RQ3}
The main results of our comparison are displayed in Table~\ref{tab:TART}. 
Based on the displayed results we can make four observations:
\begin{enumerate*}[label=(\arabic*)]
\item
The average methods (Avg and T-Avg) perform considerably worse than all other methods.
Clearly, matrix factorization is preferable over averaging baselines. %
\item
The time-based methods outperform their static counterparts by substantial margins:
TMF $\succ$ MF, 
TMF-StaticIPS $\succ$ MF-StaticIPS, and
TMF-DANCER $\succ$ MF-DANCER,
except T-Avg $\prec$ Avg due to sparsity.
This shows that assuming static preferences can substantially hurt the performance of a method when user preferences are actually dynamic.
\item
The debiased methods increase performance: MF-DANCER $\succ$ MF and \ac{TMF}-DANCER $\succ$ TMF-StaticIPS $\succ$ TMF. 
There is a single exception: \ac{MF} $\succ$ \ac{MF}-staticIPS under the assumption of static bias.
This surprising observation shows that \ac{DANCER} is more robust to certain dynamic scenarios.
\item
Finally, the best performing method is TMF-DANCER, which both models dynamic preferences and is debiased under the assumption of dynamic selection bias.
While it is not a surprise that this method performs well in the scenario that it assumes, 
the differences with other methods are considerable and statistically significant.
\end{enumerate*}

In addition, Table~\ref{tab:diff-est-prop} displays the performance of TMF-DANCER using different propensities.
We see that with estimated propensities the performance of TMF-DANCER is comparable to when it is using the actual \ac{sim-GT} propensities.
Moreover, TMF-DANCER outperforms the most baselines, except \ac{TMF}-StatisIPS, even when using simple time-aware propensity estimation.

\begin{table}[tbp]
	\centering
        \caption{\fontdimen2\font=0.32ex RQ3 -- Performance of \ac{TMF}-\ac{DANCER} compared with different methods. 
		$\dag$ indicates that the improvement of \ac{TMF}-\ac{DANCER} over all the baselines is significant at the level of 0.01. 
		} %
		\label{tab:TART}%
			\begin{tabular}{@{}l l@{~}l l@{~}l l@{~}l@{}}
			\toprule
			Method & MSE$\downarrow$ & & MAE$\downarrow$ & & ACC$\uparrow $ & \\
			\midrule
			Avg & 0.3155 & & 0.4321 & & 0.3623 & \\
			T-Avg & 0.3280 & & 0.4326 & & 0.3614 \\
			MF & 0.1811 & {\scriptsize(0.0030)} & 0.3314 & {\scriptsize(0.0028)} & 0.4680 & {\scriptsize(0.0040)} \\
            TMF & 0.1338 & {\scriptsize(0.0019)} & 0.2818 & {\scriptsize(0.0022)} & 0.5396 & {\scriptsize(0.0038)} \\
			\midrule
			MF-StaticIPS & 0.1879 & {\scriptsize(0.0035)} & 0.3377 & {\scriptsize(0.0032)} & 0.4598 & {\scriptsize(0.0044)} \\
			TMF-StaticIPS & 0.1086 & {\scriptsize(0.0021)} & 0.2491 & {\scriptsize(0.0027)} & 0.6065 & {\scriptsize(0.0057)} \\
			\midrule
			MF-\ac{DANCER} & 0.1533 & {\scriptsize(0.0016)} & 0.3032 & {\scriptsize(0.0017)} & 0.5074 & {\scriptsize(0.0023)} \\
            TMF-\ac{DANCER} & \textbf{0.1045}$^\dag$ & {\scriptsize(0.0014)} & \textbf{0.2444}$^\dag$ & {\scriptsize(0.0018)} & \textbf{0.6151}$^\dag$ & {\scriptsize(0.0039)} \\
			\bottomrule
		\end{tabular}%
\end{table}%

\begin{table}[tbp]
	\centering
	\vspace{-3mm}
        \caption{RQ3 -- Performance of \ac{TMF}-\ac{DANCER} with estimated propensities and the (simulated) ground truth propensities.}
		\label{tab:diff-est-prop}%
			\begin{tabular}{l l@{~}l l@{~}l l@{~}l}
			\toprule
			Method & MSE$\downarrow$ & & MAE$\downarrow$ & & ACC$\uparrow $ & \\
			\midrule
			TG-Pop & 0.1182 & {\scriptsize(0.0012)} & 0.2644 & {\scriptsize(0.0016)} & 0.5677 & {\scriptsize(0.0032)} \\
			T-Pop & \textbf{0.1041} & {\scriptsize(0.0015)} & 0.2448 & {\scriptsize(0.0022)} & 0.6115 & {\scriptsize(0.0055)} \\
			Ground Truth & 0.1045 & {\scriptsize(0.0014)} & \textbf{0.2444} & {\scriptsize(0.0018)} & \textbf{0.6151} &{\scriptsize(0.0039)} \\
			\bottomrule
		\end{tabular}%
\end{table}%

\begin{figure}[t]
    \centering
    \subfigure{
	  \includegraphics[width=\linewidth]{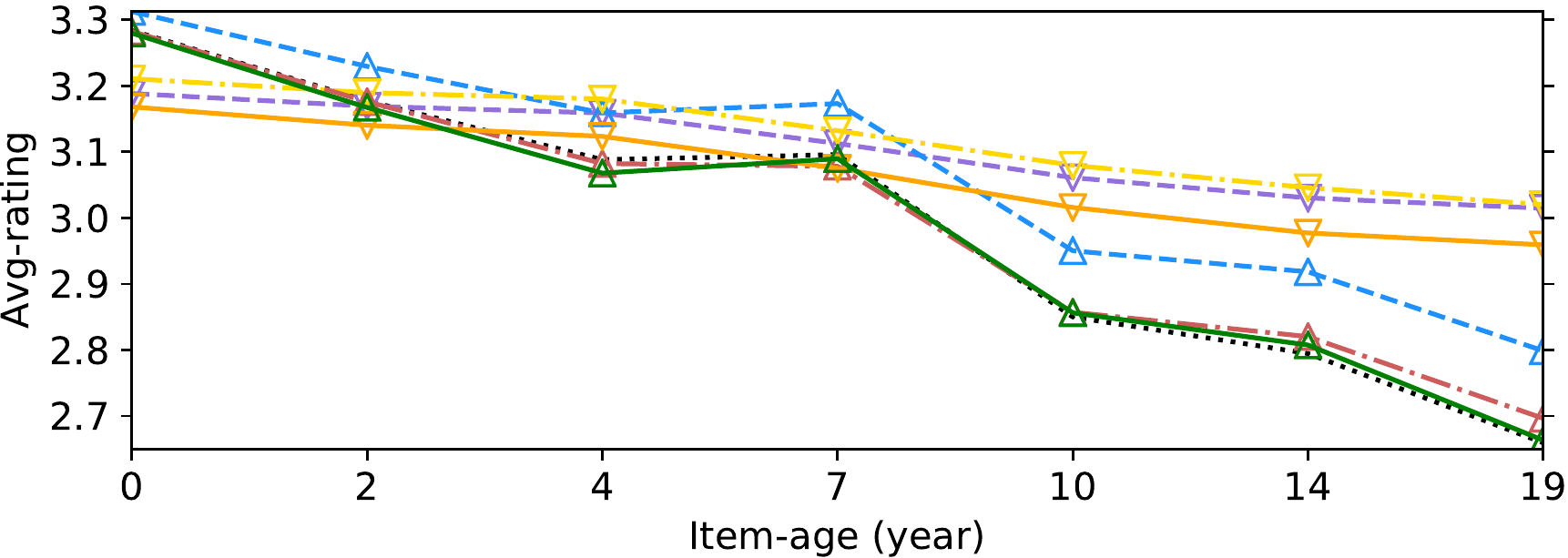}
	  } \vskip -4pt
    \subfigure{
      \includegraphics[width=\linewidth]{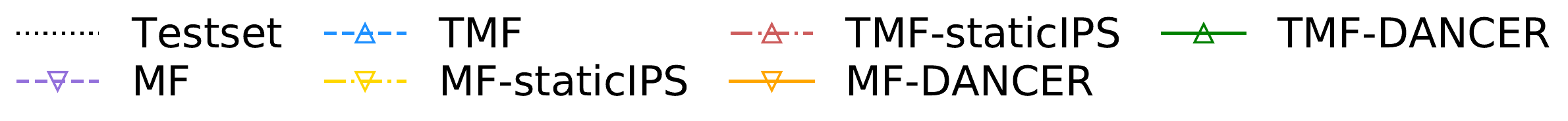}} 
	\vspace{-4mm}
    \caption{RQ3 -- Average rating on items predicted by different models over the item-age.}
    \label{fig:Model-avg-rating}
\end{figure}

To better understand the improvements of TMF-DANCER, Figure~\ref{fig:Model-avg-rating} shows the average predicted rating from different methods across item-ages and the actual average rating.
The MF methods are unable to model changes in ratings as items get older; the differences in the average ratings are purely caused by different item distributions: \ie items that become available later in the dataset will never achieve the oldest item-ages.
The TMF methods better capture the overall trend.
TMF without debiasing consistently overestimates ratings; TMF-staticIPS reduces overestimation by correcting for static bias; the overestimation becomes worse for older items in both models.
Instead, TMF-DANCER approximates the actual average rating at each item-age; its accuracy is quite consistent over time.

\begin{figure}[t]
\centering
	\vspace{-2mm}
	\bf \small Mad Max (1979)\\
	  \includegraphics[width=0.50\linewidth]{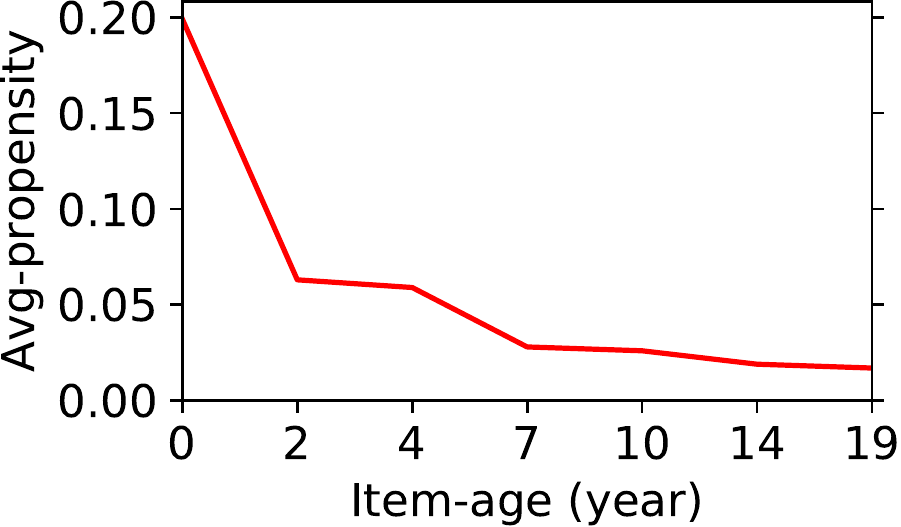}%
      \includegraphics[width=0.50\linewidth]{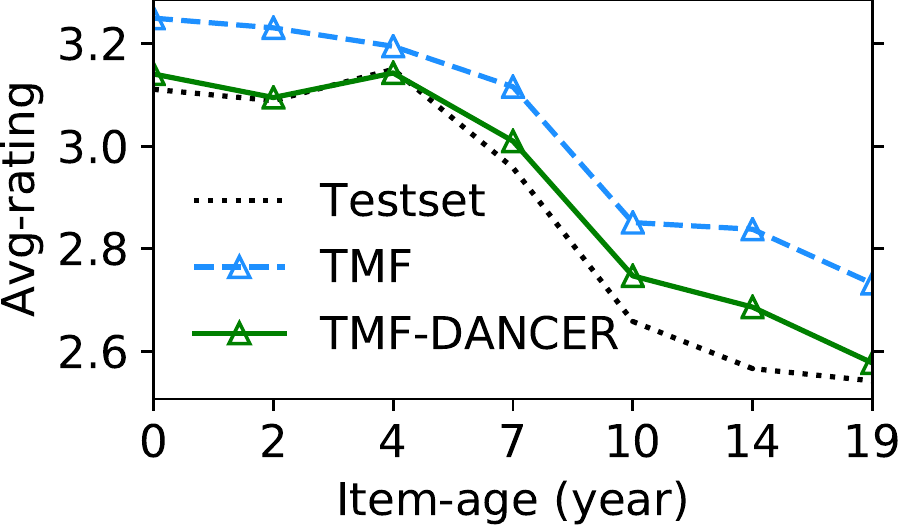}
	\\
	\bf \small Kid in King Arthur's Court (1995)\\
	  \includegraphics[width=0.50\linewidth]{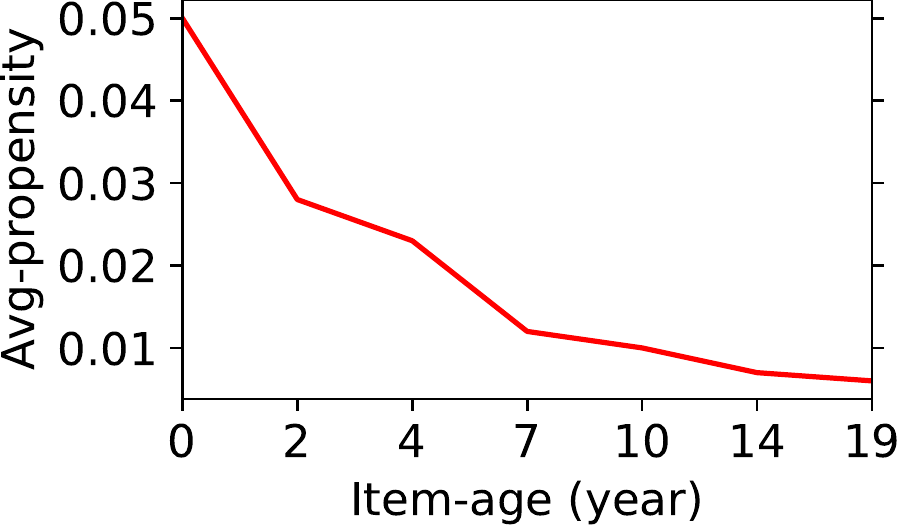}%
      \includegraphics[width=0.50\linewidth]{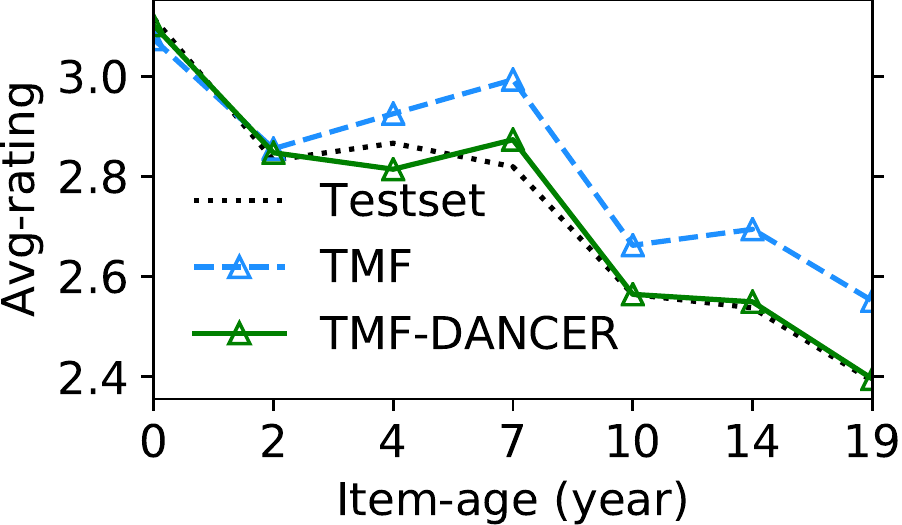}
        \caption{\fontdimen2\font=0.32ex  RQ3 -- Average propensities and predicted average rating over item-age of the very popular movie ``Mad Max (1979)'' and the less popular ``Kid in King Arthur's Court (1995)''.}
	\label{fig:examplemovies} 
\end{figure}

Lastly, to get more insights into the behavior of TMF-DANCER,
Figure~\ref{fig:examplemovies} shows the propensities and (predicted) ratings per item-age and averaged across users for two handpicked movies. 
We observe that \ac{TMF}-\ac{DANCER} outperforms \ac{TMF}, especially when the popularity of items decreases as items get older.

Finally, we can answer \textbf{RQ3} in the affirmative:
the \ac{TMF}-\ac{DANCER} method better mitigates the effect of bias in a dynamic scenario than existing debiasing methods designed for static selection bias.
This conclusion still holds when propensities are estimated, and the accuracy of \ac{TMF}-\ac{DANCER} is consistent across item-ages.

\vspace*{-1mm}
\section{Conclusion}
In this paper, we considered the dynamic scenario in recommendation where selection bias and user preferences change over time.
Our experimental results revealed that in the real-world MovieLens dataset:
\begin{enumerate*}[label=(\arabic*)]
    \item selection bias changes as items get older, and
    \item user preferences are also affected by the age of items.
\end{enumerate*}
Therefore, it appears that the dynamic scenario better captures the real-world situation, and thus, poses a serious problem that existing static IPS-based method cannot correct for dynamic bias in dynamic scenarios.
As a solution, we proposed the DANCER debiasing method that takes into account the dynamic aspects of bias and user preferences, the first method that is unbiased in the dynamic scenario.
The results on a semi-synthetic simulation based on the MovieLens dataset showed that TMF-DANCER provides significant gains in performance that are consistent across item-ages and robust to misspecified propensities.
Our findings about the dynamic scenario have implications for state-of-the-art recommendation methods, as they are strongly affected by dynamic selection bias.
With the \ac{DANCER} debiasing method, \acp{RS} can now be expanded to deal with dynamic scenarios.

A limitation of our work is that we only considered the rating prediction task and the effect of item-age on bias and preferences;
future work should consider the ranking task and look at other aspects of time, \eg seasonal effects, weekday, time of day, etc.

\begin{acks}
This research was partially supported by
the Netherlands Organisation for Scientific Research (NWO)
under pro\-ject number
024.\-004.\-022
and the Google Research Scholar Program.
All content represents the opinion of the authors, which is not necessarily shared or endorsed by their respective employers and/or sponsors.
\end{acks}
\clearpage
\bibliographystyle{ACM-Reference-Format}
\balance
\bibliography{huang21}

\end{document}